\documentclass[aps,prb,twocolumn,showpacs,superscriptaddress]{revtex4-1}
\usepackage{enumerate}
\usepackage{graphicx}
\usepackage{amsmath}
\usepackage{amsfonts}
\usepackage{amssymb}
\usepackage{epstopdf}
\usepackage{color}
\usepackage{bm} 
\newcommand{\ket}[1]{|#1\rangle}
\newcommand{\bra}[1]{\langle #1|}

\begin{document}
\today
\title{Spin-electric Berry phase shift in triangular molecular magnets}
\author{Vahid Azimi Mousolou\footnote{Electronic address: v.azimi@sci.ui.ac.ir}}
\affiliation{Department of Mathematics, Faculty of Science, 
University of Isfahan, Box 81745-163 Isfahan, Iran}
\author{C.~M.~Canali\footnote{Electronic address: carlo.canali@lnu.se}}
\affiliation{Department of Physics and Electrical Engineering, Linnaeus University,
SE-391 82 Kalmar, Sweden}
\author{Erik Sj\"oqvist\footnote{Electronic address: erik.sjoqvist@physics.uu.se}}
\affiliation{Department of Physics and Astronomy, Uppsala University, Box 516, 
SE-751 20 Uppsala, Sweden}
\begin{abstract}
We propose a Berry phase effect on the chiral degrees of freedom of a triangular 
magnetic molecule. The phase is induced by adiabatically varying an external 
electric field in the plane of the molecule via a spin-electric coupling mechanism 
present in these frustrated magnetic molecules. The Berry phase effect depends on spin-orbit 
interaction splitting and on the electric dipole moment. By varying the amplitude of 
the applied electric field, the Berry phase difference between the two spin states can 
take any arbitrary value between zero and $\pi$, which can be measured as a phase shift 
between the two chiral states by using spin-echo techniques. Our result can be used to realize 
an electric field induced geometric phase-shift gate acting on a chiral qubit encoded in 
the ground state manifold of the triangular magnetic molecule.
\end{abstract}
\pacs{03.65.Vf, 75.50.Xx, 75.75-c, 03.67.Lx}
\maketitle
\section{Introduction}
Berry's phase, originally discovered \cite{berry84} for a nondegenerate pure quantum state 
evolving adiabatically in a cyclic fashion, has been subsequently extended to nonadiabatic 
evolution \cite{aharonov87,anandan88}, the evolution of degenerate quantum states \cite{wilczek84}, 
and mixed states\cite{uhlmann86,sjoqvist00a}. In parallel to the theoretical development, these 
phases have been demonstrated experimentally in a wide variety of contexts, including optical 
\cite{tomita86}, molecular \cite{vonbusch98}, and solid-state \cite{zhang05} systems.  
The properties of the Berry phase make it an essential unifying concept in the 
physical sciences \cite{xiao10}.

The emerging and partly overlapping fields of molecular 
quantum spintronics\cite{Bogani2008} and magnon spintronics\cite{Chumak2015}
offer a promising approach to design 
devices for information storage, transport and processing.
In this development, magnetic molecules (MMs) 
\cite{gatteschi06} play a central role. MMs possess rich quantum properties, which can 
be chemically engineered. There has been considerable recent interest in MMs since they all 
have, in contrast to, e.g., nanoparticles, identical properties, which is an important advantage 
for the realization of scalable ensembles of quantum computation entities. Antiferromagnetic 
triangular molecules such as \{Cu{$_3$}\} complexes 
(e.g., Na$_{12}$[Cu$_3$(AsW$_9$O$_{33}$)$_2$ $\cdot$ 3H$_2$O] $\cdot$32 H$_2$O) \cite{choi06} 
are a special class of MMs particularly suitable for quantum 
control and manipulation. Due to the lack of inversion symmetry, 
these triangular MMs display an effective 
spin-electric coupling mechanism \cite{bulaevskii08,trif08,trif10} acting within their 
quasi-degenerate chiral ground state. This provides a proper and applicable way to do quantum 
information processing with spin systems, since electric fields are simpler to apply and control 
at small spatial scales and short time scales, compared to magnetic fields\cite{trif08,trif10}. 
For these reasons, electrical control of electron and nuclear spin qubits is a subject intensively 
investigated, not only in MMs \cite{Thiele2014}, but also in semiconductor quantum dots 
containing single electrons \cite{Kawakami2014}, and in devices consisting of single impurities 
in a semiconductor host, e.g., P-donors in Si \cite{Laucht2015}, as originally proposed by 
Kane \cite{Kane1998}.

Laser-induced nonadiabatic (femtosecond) spin dynamics in isolated triangular 
clusters, such as Co$^+_3$(EtOh), Co$^+_3$(EtOh), Ni$_3$(CH$_3$OH), and Co$^+$(CH$_3$OH) have 
been studied recently both theoretically (using advanced {\it ab-initio} methods) and experimentally 
in molecular beam experiments \cite{Yin2014, Chaudhuri2015}. These tri-nuclear transition-metal
clusters are different from the antiferromagnetic triangular molecules considered in the present 
paper. Their spin properties make them closely resemble single-molecule magnets (SMM) 
\cite{remarkSMM}. Nevertheless these studies show that efficient electric manipulations of 
the molecular spin density, in this case mediated by spin-orbit interaction, is feasible. In 
principle, similar molecular beam experiments could be extended also to triangular 
antiferromagnetic molecules such as Cu$_3$, where a  coupling between spin chiral 
states and the electric field is present even in the absence of spin-orbit coupling.

In this paper we show that in a triangular antiferromagnetic MM subject to a time-dependent 
external electric field, the spin-electric coupling induces a Berry phase in the spin-chiral ground 
state manifold. We use applied electric field pulses in the presence of a static magnetic field to 
realize a conditional dynamics of the system. We show that Berry phases with arbitrary values 
between zero to $\pi$ on the chiral degree of freedom (chiral qubit) can be achieved by adiabatically 
varying the electric field in the plane of the molecule, and can be measured by using the 
spin-echo technique. The Berry phase shift depends on the effective spin-orbit interaction and 
electric dipole moment of the MM, which are the two most important quantities that control 
the spin-electric coupling mechanism. In the adiabatic limit, the ratio of these two quantities 
can be determined by measuring the Berry phase shift as a function of the electric field amplitude. 
The Berry phase can be used to implement a phase-shift gate, with an arbitrary phase, acting 
on the chiral part of the ground state manifold of the triangular MM, which encodes a single 
chiral qubit. These gates are electric field generated and geometric, two key ingredients to 
realize flexible and coherent switching, which is needed for the implementation of efficient
quantum processors.

The outline of the paper is as follows. In the next section, we provide a brief introduction 
to spin-electric coupling in triangular MMs and their ground state manifold. In 
Sec.~\ref{Dynamics of the system}, we describe the conditional dynamics with respect 
to the spin-chirality decomposition of the system. We demonstrate how it can be used 
to test the geometric Berry phase effect on the chirality in Sec.~\ref{two-qubit 
geometric phase shift gate}. The paper is summarized in Sec.~\ref{summary}.

\section{Effective qubit system}
\label{sec:effective qubit system}
\begin{figure}[t]
\centering
\includegraphics[width=50mm,height=45mm]{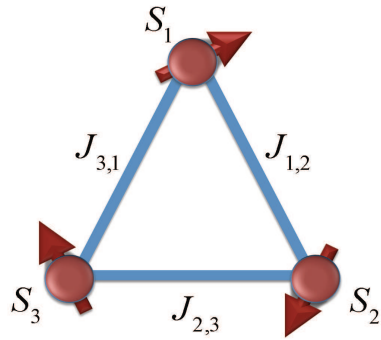}
\caption{Schematic geometry of triangular molecular magnets.}
\label{fig:TSMM}
\end{figure}

The degeneracy of the ground state (GS) of spin rings containing an odd number of
antiferromagnetically coupled half-integer spins make them a suitable candidate for 
quantum information processing. In particular, odd-number rings of half-odd integer 
spins satisfy the conditions, which allow for spin state manipulation via pulsed electric 
fields \cite{trif10}. The simplest nontrivial class of such a spin system is a triangular ring 
of $s= \frac{1}{2}$ spins, e.g., $\textrm{Cu}_3, \textrm{V}_{15}, \textrm{Co}_3$. As depicted 
in Fig.~\ref{fig:TSMM}, the magnetic core of such MMs consists of three $s= \frac{1}{2}$ 
spins positioned at the vertices of an equilateral triangle and coupled by an antiferromagnetic 
Heisenberg exchange interaction. In the absence of external fields, the system can be described 
by the spin Hamiltonian \cite{trif10}
\begin{eqnarray}
H & = & \sum_{k=1}^3 
J_{k,k+1} {\bf s}_{k}\cdot{\bf s}_{k+1} 
\nonumber \\ 
 & & + \sum_{k=1}^3{\bf D}_{k,k+1}\cdot\left({\bf s}_k \times {\bf s}_{k+1}\right), 
\label{hamiltonian2}
\end{eqnarray}
with a periodic boundary condition, where the first and fourth sites are identified. Here, ${\bf s}_k$ 
is a spin$-\frac{1}{2}$ vector operator localized at site $k$. In Eq.~(\ref{hamiltonian2}), the first 
term is an isotropic Heisenberg interaction with antiferromagnetic exchange couplings $J_{k,k+1}$, 
and the second term is an antisymmetric Dzyaloshinsky-Moriya (DM) interaction 
\cite{dzyaloshinsky58,moriya60}. Under the assumption, that the magnetic Hamiltonian 
of the molecule is invariant under point group $D_{3h}$, its parameters satisfy the constraints  
\begin{eqnarray}
J_{k,k+1}=J, \ \ {\bf D}_{k,k+1}= (0, 0,  D^{z}), \ \ k=1,2,3.
\label{d3h_conditions}
\end{eqnarray}

The GS manifold of these frustrated MMs is given by two degenerate total spin $S= \frac{1}{2}$ 
doublets ($S_{z} = \pm  \frac{1}{2}$) of opposite spin chirality $\chi = \pm 1$. Explicitly, the 
GS-space is spanned by the following linearly independent quadruplet, 
\begin{eqnarray}
\ket{\pm 1, + \frac{1}{2}} & = & \frac{1}{\sqrt{3}}[\ket{\downarrow\uparrow\uparrow}+\eta_{\pm}\ket{\uparrow\downarrow\uparrow} + 
\eta_{\mp}\ket{\uparrow\uparrow\downarrow}], 
\nonumber\\
\ket{\pm 1, - \frac{1}{2}} & = & \frac{1}{\sqrt{3}}[\ket{\uparrow\downarrow\downarrow} +
\eta_{\pm}\ket{\downarrow\uparrow\downarrow} +
\eta_{\mp}\ket{\downarrow\downarrow\uparrow}],
\nonumber\\
\eta_{\pm} & = & e^{\pm i2\pi/3},
\label{GS}
\end{eqnarray}
constructed by symmetry adapted linear combination of various possible spin configurations. 
The quantum basis states given in Eq.~(\ref{GS}) are simultaneous eigenvectors of the 
$z$-component of chirality and total spin operators,  i.e., $C_z$ and $S_z$, respectively, where 
the components of the chirality vector operator read 
\begin{eqnarray}
C_{x}&=&-\frac{2}{3}({\bf s}_1 \cdot{\bf s}_2 -2{\bf s}_2 \cdot{\bf s}_3 + {\bf s}_3 \cdot{\bf s}_1),\nonumber\\
C_{y}&=&\frac{2}{\sqrt{3}}({\bf s}_1 \cdot{\bf s}_2 - {\bf s}_3 \cdot{\bf s}_1),\nonumber\\
C_{z}&=&\frac{4}{\sqrt{3}}{\bf s}_1 \cdot({\bf s}_2 \times{\bf s}_3 ).
\label{chiral_vector}
\end{eqnarray}
One can verify that the chiral operators define the same algebra as the spin-half operators. 
Namely, $C_{k},\ k=x,y,z$, in  the chiral basis states $\ket{\chi = \pm 1}$ are the same as 
Pauli matrix components, and thus $[C_k, C_l]=2i\sum_m \epsilon_{klm}C_m$, with 
$\epsilon_{klm}$ being the Levi-Civita symbol.

The energy gap $\Delta_J$ between the GS manifold and the first excited state, the $S=\frac{3}{2}$ 
quadruplet, is typically of the order of 
1 meV.\cite{remark1}
The spin-orbit induced DM interaction lifts the 
degeneracy between the two chiral doublets with a splitting $\Delta_{\textrm{SO}} \le 0.1  
\Delta_J$ \cite{trif08,Nossa12}. What makes these triangular MMs interesting for quantum 
manipulation is that an electric field in the $xy$-plane of the molecule couples the two GS 
doublets of opposite chirality, due to the lack of inversion symmetry \cite{trif10,trif08,Islam10}. 

In the presence of external electric ($\textbf{E}$) and magnetic ($\textbf{B}$) fields, the 
dynamics  of the GS-space spanned by the basis states given in Eq.~(\ref{GS}) is described 
by the effective low-energy spin Hamiltonian \cite{trif08}
\begin{eqnarray}
H_{\textrm{eff}} = \Delta_{\textrm{SO}} C_z S_z +
p\;\textbf{E}' \cdot \textbf{C}_{\parallel} +
\textbf{B} \bar{\bar{g}} \cdot \textbf{S} 
\label{eff. hamiltonian}
\end{eqnarray}
with $\textbf{C}_{\parallel} = (C_x, C_y,0)$ and $\textbf{S}=(S_x, S_y, S_z)$ being
the chirality and spin vector operators, respectively. $\textbf{E}'=\mathcal{R}_{z}(\alpha)\textbf{E}$ 
is the electric field $\textbf{E}$ rotated about the $z$ axis by an angle $\alpha=7\pi/6-2\beta$ 
with $\beta$ being the angle between the in-plane component of the electric field $\textbf{E}$ 
and a vector pointing from site 1 to 2. Due to the symmetry of the molecule 
$\bar{\bar{g}}=\textrm{diag}\{g_{\parallel}, g_{\parallel}, g_{\bot}\}$. The parameter $p$ 
has the units of an electric dipole moment, and it gives the strength of the effective coupling 
between the two states with opposite chirality brought about by the electric field. In Cu$_3$ 
MM $p$ is not small \cite{Islam10}, and for typical electric fields generated by a scanning tunneling 
microscope (STM) ($\approx 10^2$ kV/cm) the spin-chirality manipulation (Rabi) time is 
$10-10^3$ ps \cite{trif08,Islam10}.

We conclude this section with a discussion on the validity of the effective spin Hamiltonian 
given in Eq.~(\ref{hamiltonian2}), satisfying the constraints of Eq.~(\ref{d3h_conditions}) 
imposed by the $D_{3h}$ symmetry of the molecule. It is known\cite{murao1974} that 
equilateral triangular molecules with an odd number of electrons undergo a Jahn-Teller (JT) 
distortion that reduces the $D_{3h}$ symmetry. Typically the deformation makes one of the 
sides slightly shorter or longer, leading to an isosceles triangle with $D_{2v}$ symmetry. 
For Cu$_3$ complexes and other triangular molecules the deformation is found to be very 
tiny both experimentally\cite{choi06} and theoretically\cite{Islam2016} (the side change is 
of the order of 0.001 {\AA} for Cu$_3$ complexes\cite{choi06}), and it is usually neglected 
\cite{trif08}. The JT distortion mechanism lifts the chiral degeneracy of the ground state, 
even in the absence of spin-orbit coupling. We can describe this effect within the spin Hamiltonian 
approach by adding to Eq.~(\ref{hamiltonian2}) the JT-induced correction
\begin{equation}
\delta H_{\rm JT} = 
\sum_{k=1}^3 
\delta J_{k,k+1} {\bf s}_{k}\cdot{\bf s}_{k+1} \, ,
\label{ham_JT}
\end{equation}
where $\delta J_{k, k+1}$ are the modifications in the exchange constants caused the 
changes in the bond lengths. Since the deformation is tiny, we typically have 
$\delta J_{k, k+1}/J \ll 1$, where $J$ is the coupling constant for the equilateral 
nuclear configuration. For example, for the case of Cu$_3$, where $J \ \approx $ 1 meV, 
$\delta J_{k, k+1} < $ 0.1 meV. For a distortion down to an isosceles triangle, two of 
these $\delta J_{k, k+1}$ are equal. Since $\delta H_{\rm JT}$ is still rotationally invariant 
in spin space, the total spin of the total Hamiltonian remains a good quantum number.
Therefore, $\delta H_{\rm JT}$ couples the two $S = \frac{1}{2}$ chiral GS states, lifting the 
degeneracy of the unperturbed Hamiltonian, but does not couple these to the $S=\frac{3}{2}$ 
excited state quadruplet. Using the definition of the chiral vector operator in 
Eq.~(\ref{chiral_vector}), one can see that $\delta H_{\rm JT}$ can be formally 
rewritten as
\begin{equation}
\delta H_{\rm JT} = p {\bf E}_{\rm JT}\cdot {\bf C}_{\parallel}\; ,
\label{JT_chiral}
\end{equation}
where ${\bf E}_{\rm JT}$ is an internal electric field describing the JT deformation of the 
molecule\cite{trif08}. Here the components of the vector $p {\bf E}_{\rm JT}$ can be easily 
related to the parameters $\delta J_{k,k+1}$.

Since the JT reduces the symmetry of the molecule to $D_{2v}$, the DM term is also 
modified by the JT distortion. This effect can also be studied by adding an appropriate 
spin Hamiltonian to  Eq.~(\ref{hamiltonian2}). This perturbation in general breaks rotation
symmetry in spin space, and besides coupling the two GS $S=\frac{1}{2}$ chiral states with each 
other, also couples these to the $S=\frac{3}{2}$ excited state. However, since the DM exchange 
constant for the unperturbed system is at least one-order of magnitude smaller than isotropic 
$J$, its change induced by the $JT$ distortion is typically small with respect to 
$\delta J_{k, k+1}$. Therefore, we expect the effect of this JT-induced perturbation on the
low-energy levels of the system to be considerably smaller than the one caused by 
Eq.~(\ref{JT_chiral}).

In conclusion, the main effect of the JT distortion can be described by the presence of a 
small intrinsic static in-plane electric field, which is combined to the applied external 
electric field $\bf E'$ in Eq.~(\ref{eff. hamiltonian}).

Note finally that for an isolated triangular molecule in the gas phase, there are three equivalent 
JT distorted (isosceles) configurations, all with the same GS energy but separated by an energy 
barrier. It is then possible for the system to quantum tunnel from any one of these configurations 
to the other two. This phenomenon, known as the dynamical JT effect, can effectively restore 
the original $D_{3h}$ symmetry of the molecule \cite{ham87}, 
provided that the energy barrier separating the 
three JT distorted states is small compared to perturbations that couple them.

\section{Conditional dynamics of the system}
\label{Dynamics of the system}
\begin{figure}[t]
\centering
\includegraphics[width=73mm,height=45mm]{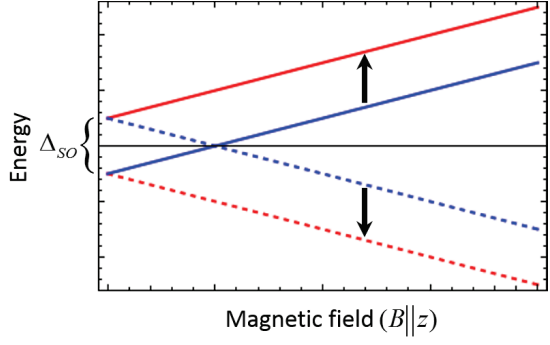}
\caption{(Color online) Schematic 
diagram of electric-field-induced transitions between states of opposite 
chirality in the ground-state manifold of a triangular MM, with the zero-field splitting 
$\Delta_{{\textrm{SO}}}$ due to the Dzyaloshinsky-Moriya interaction. Red solid (dashed) lines 
represent the states with $\chi=+1$ and $S_{z}=+ \frac{1}{2}(-\frac{1}{2})$, and blue solid 
(dashed) lines represent the states with $\chi=-1$ and $S_{z}=+\frac{1}{2}(-\frac{1}{2})$.}
\label{fig:E-SMM}
\end{figure}

In the presence of a static magnetic field in the $z$-direction, an electric field $\textbf{E}$ 
induces transitions only within each eigensubspace of $S_{z}$ of the chiral state manifold.
This implies that the effective spin Hamiltonian in Eq.~(\ref{eff. hamiltonian}) can be decomposed 
into two parts corresponding to the eigenvalues $\pm \frac{1}{2}$ of $S_{z}$. Therefore, the 
system describes two independent chiral components, decoupled from each other and split by the 
external magnetic field (see Fig.~\ref{fig:E-SMM}). For a time-dependent oscillating electric field $\textbf{E}(t)$, the 
corresponding Hamiltonian takes the form
\begin{eqnarray}
H_{\textrm{eff}}(t) & = & \frac{\hbar}{2}[\Omega_{0}I+\boldsymbol{\Omega}_{+}(t) 
\cdot \boldsymbol{\sigma}]\otimes\ket{+ \frac{1}{2}}\bra{+ \frac{1}{2}}
\nonumber\\
 & & +\frac{\hbar}{2}[-\Omega_{0}I+\boldsymbol{\Omega}_{-}(t)\cdot \boldsymbol{\sigma}] 
\otimes\ket{- \frac{1}{2}}\bra{- \frac{1}{2}} , 
\label{eq:T-effH}
\end{eqnarray}
where $\hbar\Omega_{0}=g_{\bot}B_{z}$,  
\begin{eqnarray}
\hbar\boldsymbol{\Omega}_{\pm} (t) = 
(p\mathcal{E}\cos(\omega t+\phi), p\mathcal{E}\sin(\omega t+\phi), \pm\Delta_{{\textrm{SO}}})
\end{eqnarray}
are the Rabi vectors, and $\boldsymbol{\sigma}=(\sigma_{x}, \sigma_{y}, \sigma_{z})$ is the 
Pauli vector operator that acts on the chiral degrees of freedom. $\omega$, $\phi$, and 
$\mathcal{E}$ are the angular frequency, phase, and amplitude of the oscillating electric field, 
respectively, and $\pm\Delta_{{\textrm{SO}}}$ are the zero-field energy splittings between the chiral states, 
when the spin is in state $\ket{\pm \frac{1}{2}}$.

The chirality dynamics conditioned on the spin state is described by 
\begin{eqnarray}
\frac{d}{dt}{\bf s}_{\pm}(t) = \boldsymbol{\Omega}_{\pm}(t)\times{\bf s}_{\pm}(t),
\label{eq:T-evolution}
\end{eqnarray}
where ${\bf s}_{\pm}(t)$ are the instantaneous Bloch vectors parametrizing the conditional 
chiral density operators 
\begin{eqnarray}
\rho_{\pm}(t)=\frac{1}{2}[I+{\bf s}_{\pm}(t)\cdot \boldsymbol{\sigma}]\otimes\ket{\pm 
\frac{1}{2}}\bra{\pm \frac{1}{2}} , 
\label{eq:ccdo}
\end{eqnarray}
given by solving the equations of motion 
\begin{eqnarray}
i\hbar\frac{d}{dt}\rho_{\pm}(t)=[H_{\textrm{eff}}(t), \rho_{\pm}(t)]. 
\end{eqnarray}
In the rotating frame with angular frequency $\omega$ around the $z$-axis, 
Eq.~(\ref{eq:T-evolution}) is equivalent to 
\begin{eqnarray}
\frac{d}{dt}{\bf s}_{\pm}^{\prime}(t) = \boldsymbol{\Omega}^{\prime}_{\pm} 
\times {\bf s}_{\pm}^{\prime}(t)
\label{eq:T-evolution1}
\end{eqnarray}
with the time-independent Rabi vectors 
\begin{eqnarray}
\hbar\boldsymbol{\Omega}^{\prime}_{\pm}=(p\mathcal{E}\cos\phi, p\mathcal{E}\sin\phi, \pm\Delta_{{\textrm{SO}}}-\hbar\omega).
\end{eqnarray}
If the Bloch vector ${\bf s}_{+}^{\prime}$ (${\bf s}_{-}^{\prime}$) is initially aligned with 
$\boldsymbol{\Omega}^{\prime}_{+}$ ($\boldsymbol{\Omega}^{\prime}_{-}$), it remains 
aligned with $\boldsymbol{\Omega}^{\prime}_{+}(\boldsymbol{\Omega}^{\prime}_{-})$ 
under an adiabatic variation of the electric field parameters $\mathcal{E}$ and $\phi$. Therefore, 
since we can realize different vectors $\boldsymbol{\Omega}^{\prime}_{+}$ 
($\boldsymbol{\Omega}^{\prime}_{-}$) by controlling the electric field parameters, 
we can adiabatically move the Bloch vector ${\bf s}_{+}^{\prime} ({\bf s}_{-}^{\prime})$ 
into different positions on the Bloch sphere. 

\section{geometric phase shift}
\label{two-qubit geometric phase shift gate}

\begin{figure}[t]
\centering
\includegraphics[width=45mm,height=54mm]{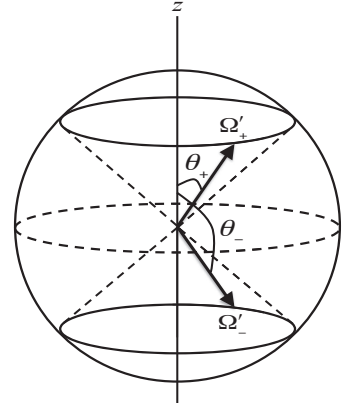}
\caption{Schematic picture of the paths in parameter space corresponding to cyclic evolutions 
of conditional chiral states in the adiabatic limit. Depending on whether the initial chiral Bloch 
vector ${\bf s}_{\pm}^{\prime}$ is parallel or anti-parallel to the initial Rabi vector 
$\boldsymbol{\Omega}^{\prime}_{\pm}$, the Bloch vector remains parallel or anti-parallel, 
respectively, to the Rabi vector along the adiabatic evolution. Thus, the polar angle 
$\tilde{\theta}_{\pm}$ between the Bloch vector ${\bf s}_{\pm}^{\prime}$ and the $z$-axis 
throughout the precession would be $\theta_{\pm}$ or $\pi-\theta_{\pm}$ depending on 
parallel condition between the initial Bloch and Rabi vectors.}
\label{fig:precession}
\end{figure}

In the absence of an electric field, the energy eigenstates of the Hamiltonian in Eq.~(\ref{eq:T-effH}) 
are the basis states given in Eq.~(\ref{GS}). Hence, the Bloch vector ${\bf s}_{\pm}^{\prime}$ 
corresponding to each energy eigenstate is either parallel or anti-parallel to 
$\boldsymbol{\Omega}^{\prime}_{\pm}$. This implies that by varying the effective Hamiltonian 
in Eq.~(\ref{eq:T-effH}) adiabatically by slowly changing the parameters of the electric field, 
we can let each energy eigenstate evolve in a cyclic fashion. Figure \ref{fig:precession} depicts 
the cyclic evolution $\mathcal{C}$ accomplished first by slowly increasing the field amplitude 
from zero to $\mathcal{E}$, then precessing the field around the $z$-axis by slowly varying the 
phase $\phi$, and finally switching the field off by slowly decreasing its amplitude to zero.

In such an adiabatic evolution, each energy eigenstate accumulates a Berry phase, being proportional 
to the solid angle subtended by its corresponding path on the Bloch sphere representing the 
chirality state space $\textrm{Span} \{ \ket{\pm 1} \}$. The associated 
geometric phases can be calculated by specifying the polar angle $\tilde{\theta}_{\pm}$ between the 
Bloch vector ${\bf s}_{\pm}^{\prime}$ and the $z$-axis throughout the precession. In the adiabatic 
regime, this angle is given by the angle between the Rabi vector 
$\boldsymbol{\Omega}^{\prime}_{\pm}$ and the $z$-axis, i.e.,
\begin{eqnarray}
\cos\theta_{\pm}=\frac{\pm\Delta_{{\textrm{SO}}}-\hbar\omega} 
{\sqrt{(\pm\Delta_{{\textrm{SO}}}-\hbar\omega)^{2}+(p\mathcal{E})^{2}}}.
\end{eqnarray}

Note that, depending on whether the initial Bloch vector is  parallel or anti-parallel to the initial 
Rabi vector, the polar angle $\tilde{\theta}_{\pm}$ is $\theta_{\pm}$ or $\pi-\theta_{\pm}$, 
respectively (see Fig.  \ref{fig:precession}). Along this evolution, the instantaneous energy 
eigenstates can be parametrized as   
\begin{eqnarray}
\ket{\Psi_{\pm}(\zeta, \xi)} & = & [\cos(\zeta/2)\ket{+1}+e^{i\xi}\sin(\zeta/2)\ket{-1}] 
\nonumber \\
 & & \otimes\ket{\pm \frac{1}{2}},
\end{eqnarray}
where $\zeta$ varies smoothly between zero and the polar angle $\tilde{\theta}_{\pm}$, and 
$\xi$ changes slowly between zero and $2\pi$. Using this parametrization, one can calculate 
the Berry phases as 
\begin{eqnarray}
\gamma_{\pm} & = & i\oint_{\mathcal{C}}\bra{\Psi_{\pm}(\zeta, \xi)}d\ket{\Psi_{\pm} 
(\zeta, \xi)}
\nonumber\\
 & = & i\int_{0}^{2\pi}\bra{\Psi_{\pm}(\tilde{\theta}_{\pm}, \xi)}\frac{d}{d\xi} 
\ket{\Psi_{\pm}(\tilde{\theta}_{\pm}, \xi)}d\xi 
\nonumber\\
 & = & -\pi(1-\cos\tilde{\theta}_{\pm}).
\label{gammapm}
\end{eqnarray}
Considering the fact that the polar angles $\tilde{\theta}_{\pm}$ depend on the orientation of 
initial Bloch vectors we obtain the following Berry phases 
\begin{eqnarray}
\gamma_{+1,+\frac{1}{2}} & = & -\gamma_{-1,+\frac{1}{2}}=\gamma_{+}=-\pi(1-\cos\theta_{+}), 
\nonumber\\
\gamma_{-1,-\frac{1}{2}} & = & -\gamma_{+1,-\frac{1}{2}}=\gamma_{-}=-\pi(1-\cos\theta_{-})
\end{eqnarray}
with corresponding dynamical phases 
\begin{eqnarray}
\delta_{\pm1,+\frac{1}{2}} & = & \frac{-1}{2\hbar}\int_{0}^{T}\bigg[g_{\bot}B_{z} 
\pm\sqrt{(\Delta_{{\textrm{SO}}}-\hbar\omega)^{2}+(2p\mathcal{E})^{2}}\bigg]dt,
\nonumber\\
\delta_{\pm1,-\frac{1}{2}} & = & \frac{1}{2\hbar}\int_{0}^{T}\bigg[g_{\bot}B_{z}\pm\sqrt{(\Delta_{{\textrm{SO}}}+\hbar\omega)^{2} +
(2p\mathcal{E})^{2}}\bigg]dt. 
\nonumber\\
\end{eqnarray}
Clearly, the cyclic evolution $\mathcal{C}$ yields the unitary phase transformation
\begin{eqnarray}
\ket{x, y}\longrightarrow e^{i(\gamma_{x, y}+\delta_{x, y})}\ket{x,y}, \ \ \ \ x, 2y =\pm 1,
\end{eqnarray}
of the spin-chirality basis vectors. 

In order to realize purely geometric phase shifts, it is necessary to eliminate the 
dynamical phases $\delta_{x, y}$. This can be achieved by using a technique known 
as {\it spin-echo} \cite{ekert00, jones00}. In this procedure, we apply the cyclic evolution 
$\mathcal{C}$ combined with fast $\pi$ transformations, which simply flip the spin or chiral 
basis states by applying in the molecular plane a pulsed magnetic or electric field, respectively. 
Explicitly, we let the system evolve through the following compound evolution
\begin{eqnarray}
C_{\text{net}}:\ \ \mathcal{C}\rightarrow \pi_{2}\rightarrow\mathcal{C}\rightarrow \pi_{1}\rightarrow\mathcal{C}^{-1}\rightarrow \pi_{2}\rightarrow\mathcal{C}^{-1}\rightarrow \pi_{1}\nonumber\\
\end{eqnarray}   
where $\pi_{2}$ ($\pi_{1}$) is a fast spin (chiral) flip transformation, and the path 
$\mathcal{C}^{-1}$ in parameter space is the same as $\mathcal{C}$ described above, 
but in the reverse direction. The net effect of this compound transformation would be 
that the dynamical phases are all canceled out and we are only left with geometric phase 
factors. Thus, the net unitary transformation is given by 
\begin{eqnarray}
U(C_{\textrm{net}}) & = & \left( e^{i2\Delta\gamma} \ket{+1} \bra{+1} \right. 
\nonumber \\ 
 & & \left. + e^{-i2\Delta\gamma} \ket{-1} \bra{-1} \right)  \otimes \hat{1}_{\textrm{spin}} , 
\label{phase-gate}
\end{eqnarray}
purely dependent on the Berry phase shift 
\begin{eqnarray} 
\Delta\gamma & = & \gamma_{+}-\gamma_{-}=\pi(\cos\theta_{+}-\cos\theta_{-}) 
\nonumber \\ 
 & = & \pi \left( \frac{\Delta_{{\textrm{SO}}}-\hbar\omega}{\sqrt{(\Delta_{{\textrm{SO}}} - 
\hbar\omega)^{2}+(p\mathcal{E})^{2}}} \right. 
\nonumber\\
 & & \left. +\frac{\Delta_{{\textrm{SO}}}+\hbar\omega}{\sqrt{(\Delta_{{\textrm{SO}}} + 
\hbar\omega)^{2}+(p\mathcal{E})^{2}}} \right) , 
\end{eqnarray}
and acting nontrivially only on the chiral degree of freedom. This follows from the fact that 
the cyclic adiabatic evolution in opposite directions induce the same dynamical phases but 
Berry phases with opposite signs. The unitary operator $U(C_{\textrm{net}})$ can be viewed 
as a geometric phase-shift gate \cite{ekert00} $\ket{\pm 1} \mapsto e^{\pm i2\Delta \gamma} 
\ket{\pm 1}$ acting on a chiral qubit encoded in the ground state manifold of the MM. 

\begin{figure}[t]
\centering
\includegraphics[width=82mm,height=53mm]{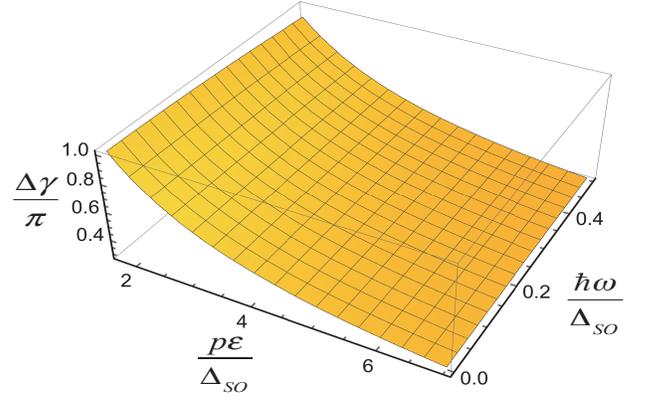}
\caption{Geometric phase shift $\Delta\gamma$ as a function of $\frac{\hbar\omega}
{\Delta_{{\textrm{SO}}}}$ 
and $\frac{p\mathcal{E}}{\Delta_{{\textrm{SO}}}}$, where $\omega$ is the angular frequency of the applied 
electric field, $p$ is the strength of the spin-electric coupling, $\mathcal{E}$ is the applied field 
amplitude, and $\Delta_{{\textrm{SO}}}$ is the zero-field energy splitting of the chiral states.}
 \label{fig:phase-shift}
\end{figure}

Figure \ref{fig:phase-shift} shows that by careful control of the electric field amplitude any 
Berry phase shift associated with the chiral qubit can be realized. It is worth noticing that 
$\Delta\gamma$ is a Lorentzian shaped phase shift with no local extrema and is independent 
of the applied external magnetic field. In the adiabatic limit $\omega\rightarrow 0$, we obtain 
\begin{eqnarray} 
\lim_{\omega\rightarrow 0}\Delta\gamma = 
2\pi\frac{\Delta_{{\textrm{SO}}}}{\sqrt{(\Delta_{{\textrm{SO}}})^{2}+(p\mathcal{E})^{2}}} ,
\label{ABP}
\end{eqnarray}
which establishes a fundamental relation between three quantities: Berry phase $\Delta\gamma$, 
the strength of the spin-electric coupling $p$, and the zero-field energy splitting $\Delta_{{\textrm{SO}}}$. 

The relation in Eq.~(\ref{ABP}) provides in principle a method for obtaining an estimate of the 
ratio $\Delta_{{\textrm{SO}}}/p$ of the two primary intrinsic quantities regulating the  spin-electric 
coupling mechanism in these MMs. By measuring the geometric Berry phase $\Delta\gamma$ 
as a function of the external electric field amplitude $\mathcal{E}$, the ratio can be evaluated. 

The geometric phase can be measured interferometrically, by canceling the dynamical phases 
picked up along different interference pathways. In the MM case, one may measure the 
Berry phase $\Delta\gamma$ by using a spin-echo interference setting, where the spin degree 
of freedom plays the role of the interferometer arms. One way to realize this would be to 
initialize the interferometer in an eigenstate of $\sigma_{x} \otimes \hat{1}_{\textrm{spin}}$ 
with, e.g., a short $\frac{\pi}{2}$ electric field pulse applied in the plane of the molecule. The 
system subsequently evolves through the spin echo compound evolution $C_{\text{net}}$ described 
above, by carefully controlling the external fields. Right after completing the spin echo sequence, another $\frac{\pi}{2}$ electric field pulse is applied 
followed by measuring the chirality of the output. One finds the probabilities 
\begin{eqnarray}
P(\textrm{chirality}=+1) & = & \cos^2 (2\Delta\gamma) , 
\nonumber \\ 
P(\textrm{chirality}=-1) & = & \sin^2 (2\Delta\gamma) 
\end{eqnarray}
from which the Berry phase shift can be extracted. A direct measurement of the chirality 
in these MMs is a non-trivial task. However, as it was pointed by Khomskii {\it et al.} 
\cite{bulaevskii08, khomskii2010}, the chiral GS states are characterized by the presence 
of spontaneous orbital currents, giving rise to magnetic orbital moments proportional to 
the chirality. Therefore it is in principle possible to determine the state chirality by carrying 
out a measure of the orbital moment via, e.g., Stern-Gerlach-type experiments.

We conclude this section with two remarks on the effect of the external magnetic field. First, 
we note that in the conditional chiral spin dynamics described above, a static magnetic field 
perpendicular to the plane of the MM plays only the passive role of splitting the two chiral 
components with opposite spin quantum numbers. Importantly, in creating ensembles of 
solid-state molecular spin qubits this function of the magnetic field is one possible solution 
to the unwanted long-range magnetic dipolar interaction, which is one of the strongest 
source of decoherence in crystals of quantum molecular magnets \cite{Takahashi2011}. 
However, a large magnetic field also renders coherent manipulations of spin qubits impractical. 
An alternative way of controlling spin decoherence of a spin qubits in a solid-state 
environment involves diluting the concentration of spin, but this has the drawback of 
weakening the nearest-neighbor spin interaction, needed for qubit entanglement. 
Quite recently a novel and more ingenious technique based on ``atomic clock transitions'' 
has been demonstrated \cite{Shiddiq2016}. In our case this is not an issue, since the 
chiral qubit of a triangular MM is entirely manipulated by the electric field, while a constant 
magnetic field is still used to freeze out one of the two spin components. 

The second remark concerns the effect of a component of a constant magnetic field in the 
plane of the molecule. When this is present, spin-up and spin-down states are coupled by 
the transverse field $B_{xy}$. Nevertheless, if $B_{xy} \ll B_z$, then the eigenstates of the 
effective Hamiltonian are still of predominant spin character, with a small admixture of the 
opposite spin contribution. When a time-dependent field is applied, we can imagine repeating 
the same analysis of the chiral state dynamics induced by the spin-electric coupling. Now, 
however, since the states are no longer pure spin state, the electric field couples chiral states 
of opposite chirality and predominately opposite spin. Using two states of opposite 
chirality and opposite spin to encode a qubit has the advantage that transitions between 
them can be more easily read out via a detection of the spin flip\cite{meier2003}.

\section{Summary}
\label{summary}
In summary, we have investigated the Berry phase effect in the system of triangular 
antiferromagnetic molecular magnets (MMs). We have demonstrated the existence of 
a Berry phase associated with the chiral degree of freedom (chiral qubit) of the system 
that can be measured by using spin-echo techniques. We show that a nontrivial Berry 
phase shift can be realized even with an in-plane external field, due to the presence 
of a spin-orbit term that couples the chiral and spin degrees of freedom. We have 
derived a unifying relation between the Berry phase, as a function of electric field and 
the two primary intrinsic quantities in the MM being the spin-orbit coupling strength and 
the electric dipole moment. In this way, the ratio between these quantities can be 
estimated by measuring the Berry phase shift on the chiral degree of freedom. Furthermore, 
by considering the two chiral states as defining a qubit embedded in the ground state 
manifold of the triangular MM, the Berry phase effect can be interpreted as a single-qubit 
geometric phase shift gate. 
The research of this paper provides an experimental testbed for exploring the physical 
nature of the Berry's phase effect in solid state systems.

We have confined ourselves to the study of the chiral dynamics of {\it isolated} triangular 
MMs. As we mentioned in the Introduction, these systems can be possibly addressed in
molecular beam experiments, similar to the ones realized in Ref.~\onlinecite{Yin2014}. 
An alternative experimental realization of the effect studied in this paper could possibly 
involve the functionalization of the MMs onto an appropriate surface/substrate, which are 
then electrically addressed by a nearby STM tip and electric gates. The choice of the surface 
is crucial. First of all the $D_{3h}$ symmetry of the molecule has to be preserved in 
such a way that the simple theoretical model discussed above is  applicable. Graphene 
and boron nitride are both substrates that display the correct crystal symmetry. Secondly, 
unwanted charge-transfer effects between the molecule and substrate that would mask and 
complicate the realization of the spin-electric coupling must be avoided or controlled. 
This is in fact a challenging task. Ongoing first-principles calculations \cite{Islam2016} 
of Cu$_3$ and V$_3$ MMs on graphene and boron nitride substrates can provide useful 
hints on the effect of the environment on the chiral GS manifold and the spin-electric 
coupling in these antiferromagnetic triangular molecular magnets.  
Another interesting possibility to realize experimentally the effects proposed in this paper 
consists in utilizing self-regulated atom trapping in open nanocorrals to built triangular 
clusters on surfaces with atomic-level precision and without the need for ligands 
\cite{Cao2014}.

\section*{Acknowledgments}
One of us (C.M.C.) would like to thank Md F. Islam and M.R. Pederson
for interesting discussions on the first-principles studies
of triangular antiferromagnetic molecules.
This work was supported by Department of Mathematics at University of Isfahan (Iran), and 
Department of Physics and Electrical Engineering at Linnaeus University (Sweden). C.M.C. 
acknowledges financial support from the Swedish Research Council (VR) through Grant No. 
621-2014-4785. E.S. acknowledges financial support from the Swedish Research Council (VR) 
through Grant No. D0413201.


\begin{thebibliography}{99}
\bibitem{remark1} 
Both available experimental studies on triangular molecules, e.g. Cu$_3$\cite{choi06}, 
and considerable computational work done within density functional theory (DFT) 
\cite{Islam10, Nossa12} seem to 
indicate that the GS is indeed  a doubly degenerate $S=\frac{1}{2}$ doublet, with degeneracy 
lifted by a spin-orbit interaction and possibly by the Jahn-Teller effect, as detailed above. 
However, the nature of the (super) exchange coupling in these molecular systems is complex,
since the effective interaction between magnetic ions is mediated via long bridges of
non-magnetic atoms. We cannot exclude that in some cases, possibly for systems with
strong correlations, effective exchange constants with ferromagnetic (FM) coupling might 
become dominant, leading to the highest multiplicity state being the GS.
\end{thebibliography}
\end{document}